\documentstyle[prc,twocolumn,aps,epsfig]{revtex}

\begin{document}
\preprint{FZJ-IKP-TH-2003-05}
\draft

\title{Charge symmetry breaking as a probe for the
real part of $\eta$--nucleus scattering lengths}

\author{V. Baru$^{1,2}$, J. Haidenbauer$^1$, C. Hanhart$^1$, and J. A.
Niskanen$^3$}

\address{$^1$Institut f\"ur Kernphysik, Forschungszentrum
J\"ulich, D-52425 J\"ulich, Germany, \\
$^2$Institute of Theoretical and Experimental Physics,
117259, B. Cheremushkinskaya 25,\\ Moscow, Russia \\
$^3$Department of Physical Sciences, PO Box 64,
FIN-00014 University of Helsinki, Finland}

\maketitle

\begin{abstract}
We demonstrate that one can use the occurrence of charge symmetry breaking
as a tool to explore the $\eta$--nucleus interaction near the
$\eta$ threshold. Based on indications that the cross section ratio
of $\pi^+$ and $\pi^0$ production on nuclei deviates from the
isotopic value in the vicinity of the $\eta$ production threshold,
due to, e.g., $\pi^0-\eta$ mixing, we argue that a systematic study of this ratio
as a function of the energy would allow to pin down
the sign of the real part of the $\eta$-nucleus scattering length.
This sign plays an important role in the context of the possible
existence of $\eta$-nucleus bound states.
\end{abstract}

\vspace{0.8cm}
\newcommand{\boldpi}{\mbox{\boldmath $\pi$}}
\newcommand{\boldtau}{\mbox{\boldmath $\tau$}}
\newcommand{\boldT}{\mbox{\boldmath $T$}}
\newcommand{\gaprox}{$ {\raisebox{-.6ex}{{$\stackrel{\textstyle >}{\sim}$}}} $}
\newcommand{\saprox}{$ {\raisebox{-.6ex}{{$\stackrel{\textstyle <}{\sim}$}}} $}

During the last decade or so the $\eta$ interaction with nucleons
and nuclei has attracted much attention both experimentally and
theoretically. One reason for this excitement is the
possibility of the formation of $\eta$-nucleus bound states.
The existence of such so-called $\eta$-mesic nuclei was first
predicted by Haider and Liu \cite{HaLi} based on the observation
that the elementary $\eta N$ interaction is attractive and
relatively strong \cite{Bahle}. It is expected that the
attraction gets increasingly stronger with increasing mass
number of the nuclei and eventually should lead to a bound
state. However, so far it is unclear for which mass number
that actually happens. For example, in the literature one can
find speculations that even the $\eta d$ system might already
form such a bound state \cite{x1} which, however, is 
disputed by other investigations \cite{x2}. 
More conservative estimations
consider the $\eta ^4{\rm He}$ system as the lightest
possible candidate \cite{wycech,willis,HaLi2}.

The occurence of a bound state near the reaction threshold will be
also reflected in the corresponding scattering length \cite{newton}.
In such a case the (real part of the) scattering length should
be relatively large and negative. (We adopt here the sign
convention of Goldberger and Watson \cite{watson}.)
Studies of the $\eta$-nucleus interaction near
threshold can be used to determine the $\eta$-nucleus scattering
length, and then, in principle, would permit conclusions on the
existence of such $\eta$-nucleus bound states. Information
on the $\eta$-nucleus interaction can be deduced from analysing
the energy dependence of $\eta$ production reactions such as
$pn \to d\eta$, $pd \to\, ^3{\rm He}\eta$, etc. But,
unfortunately, the energy dependence of the production cross section
of those reactions itself is not sensitive to the sign of the real part
of the scattering length, but only to its magnitude. Therefore, in
the present paper, we want to propose a complementary analysis that
would then also allow to constrain the sign of the $\eta$-nucleus scattering
length.

Recently, it was suggested that the study of $\pi$ production in
nucleon--nucleus and nucleus--nucleus collisions at energies around
the $\eta$ production threshold could
allow to obtain information on charge symmetry breaking effects
caused by $\pi - \eta$ mixing \cite{wilkin,machner,machner2,machner3,wycech2}.
Specifically, in Refs. \cite{machner,machner2,machner3}
the authors proposed to measure the cross section ratio
for the production of $^3{\rm H} \pi^+$ and $^3{\rm He} \pi^0$ in
$pd$ collisions, i.e. the ratio
\begin{eqnarray}
R=
\frac{d\sigma}{d\Omega}(pd \to ^3\!{\rm \!H}\pi^+) /
\frac{d\sigma}{d\Omega}(pd \to ^3\!{\rm \!He}\pi^0).
\label{rdef0}
\end{eqnarray}

Utilizing a simple phenomenological model these authors derived
the following result for the ratio $R$:
\begin{eqnarray}
R &\simeq&
\frac{p_{\pi^+}}{p_{\pi^0}}\frac{|{\cal M}_{\pi^+}|^2}
{|{\cal M}_{\tilde \pi^0} +\theta_m{\cal M}_{\eta}|^2}
\nonumber
\\
= && \hspace*{-0.3cm}
\frac{p_{\pi^+}}{p_{\pi^0}}\frac{2}{1+(2\theta_m\mbox{Re}
({\cal M}_{\eta}{\cal M}_{\tilde \pi^0}^*)+\theta_m^2 |{\cal M}_{\eta}|^2)
/|{\cal M}_{\tilde \pi^0}|^2}.
\label{rdef}
\end{eqnarray}
Here ${\cal M}_{\pi^+}$ etc. are the corresponding production amplitudes
and the tilded quantity in the denominator indicates that this is the
isospin state and not the physical state, i.e
${\cal M}_{\tilde\pi^0} \ = \ {\cal M}_{\pi^+} / \sqrt{2}$.
The quantity $\theta_m$ is the $\pi^0-\eta$ mixing angle.
Note that in Refs. \cite{machner,machner2,machner3} the term
$\theta_m^2 |{\cal M}_{\eta}|^2$ was neglected.
If isospin is conserved then the ratio $R$ should be equal to two.
However, there are indeed
experimental indications of significant deviations from
this value \cite{machner,machner2,machner3}.

The measurement of this cross section ratio at the COSY facility in
J\"ulich was suggested with the main motivation to quantify the
effects from charge symmetry breaking and even to determine the
$\pi^0-\eta$ mixing angle.
We will argue in the present paper, that the ratio $R$ defined in
Eq. (\ref{rdef}) is possibly an even more useful quantity
for something else, namely for determining the sign of
the $\eta$-nucleus scattering length, which in turn is related to
the possible existence of $\eta$-nuclear quasibound states. The basic 
observation behind this idea is that the expression on the very right 
hand side of Eq. (\ref{rdef}) should still be valid, if we drop the assumption
that the effects from charge symmetry breaking are given by $\pi^0-\eta$ mixing alone.
All we assume is, and this is crucial, that the additional piece
which causes the ratio to deviate from two
is strongly energy dependent and should be
proportional to the amplitude for $\eta$-nucleus scattering.

To be concrete, let us parameterize the $\eta$-nucleus production amplitude by
\begin{equation}
{\cal M}_{\eta} = {\cal M}^0_\eta \ T_\eta =
\frac{{\cal M}^0_\eta\, }{1-ip_\eta a(\eta A)} \ ,
\label{tmat}
\end{equation}
which takes into account the well-known fact that the energy dependence
of such production reactions is primarily determined by the interaction
of the particles in the final state \cite{watson}. In the
present case this interaction is given by the
$\eta$-nucleus scattering amplitude, $T_\eta$. The
$T$-matrix is approximated here by the lowest order term in the
effective-range expansion
where $a(\eta A)$ is the complex valued $\eta$-nucleus scattering
length and $p_\eta$ is the relative momentum of the $\eta$ with
respect to the nucleus. The constant ${\cal M}^0_\eta$ parameterizes
the overall strength of the production amplitude. For a specific
reaction the constants
${\cal M}^0_\eta$ and $a(\eta A)$ can be determined by a fit
to corresponding (near-threshold) cross section data.
However, the production of a real $\eta$ as in $pd \to\,
^3{\rm He}\eta$ is sensitive to $|{\cal M}_{\eta}|^2$ only.
As a consequence, it is not possible to pin down the sign of the
real part of the scattering length just from fitting to such data.
E.g., for that particular reaction the values
\begin{eqnarray}
\nonumber
|\mbox{Re}(a(\eta ^3{\rm He}))| &=& (3.8 \pm 0.6) \ \mbox{fm} \ , \\
\mbox{Im}(a(\eta ^3{\rm He})) &=& (1.6\pm 1.1) \ \mbox{fm} \
\label{aetaex}
\end{eqnarray}
were extracted from the data \cite{etadata}.
Also subsequent analyses of those data within theoretical models
did not yield unique results. While C. Wilkin \cite{wilkin2}
reported a negative sign for ${\rm Re}\, a(\eta A)$, based on an
optical potential approach, this was not confirmed by a more
refined study later on, using multiple scattering theory, carried
out by Wycech et al. \cite{wycech}, who arrived at positive
values.

In contrast to the total cross section for $pd \to\, ^3{\rm He}\eta$, the
ratio $R$ as defined in Eq. (\ref{rdef}) is sensitive
to Re$({\cal M}_{\eta}{\cal M}_{\tilde \pi^0}^*)$ and
consequently, as we will demonstrate below, also to the
sign of ${\rm Re}\, a(\eta A)$ and therefore it can provide
additional and independent information.
Let us write Re$({\cal M}_{\eta}{\cal M}_{\tilde \pi^0}^*)$
as
\begin{eqnarray}
\nonumber
\mbox{Re}({\cal M}_{\eta}{\cal M}_{\tilde \pi^0}^*)&=&
|{\cal M}_{\tilde \pi^0}| |{\cal M}^0_{\eta}|\\
&&\times
(\cos (\phi)\mbox{Re}(T_{\eta})+
\sin (\phi)\mbox{Im}(T_{\eta}))  .
\label{re}
\end{eqnarray}
Here $\phi$ is the phase between the amplitudes ${\cal M}_{\tilde \pi^0}$
and ${\cal M}^0_{\eta}$.
Pion production around the $\eta$ threshold involves already many partial
waves, as is obvious from a comparision of the data for different
proton-pion relative angles given in Fig. 1 of Ref.~\cite{machner}.
Thus, it is clear that
the phase $\phi$ must necessarily depend on the pion production angle.
However, and this is important, its variation with momentum
(or energy) is very slow and practically negligible compared to
the strong energy dependence induced by the $\eta$-nucleus interaction
in the vicinity of the $\eta$ production threshold.
Therefore, the energy dependence of Re$({\cal M}_{\eta}
{\cal M}_{\tilde \pi^0}^*)$ is given entirely by the energy dependence
of $T_{\eta}$. Above the $\eta$ production threshold the
$\eta$ momentum $p_\eta$ is real and thus
\begin{eqnarray}
\nonumber
\mbox{Re} ({T_{\eta}} )
&=&\frac{1+p_\eta a_I}
{1+2a_Ip_\eta+|a(\eta A)|^2p_\eta^2} \ , \\
\mbox{Im}(T_{\eta} ) &=&\frac{p_\eta a_R}
{1+2a_Ip_\eta+|a(\eta A)|^2p_\eta^2} \ ,
\end{eqnarray}
where we used $a(\eta A)=a_R+i a_I$.
Below the threshold, however, we have to use the analytic continuation
for $p_\eta = i\bar p_\eta$, where $\bar p_\eta$ is a positive real
number. Then
\begin{eqnarray}
\nonumber
\mbox{Re}(T_{\eta})&=&\frac{1+\bar p_\eta
a_R }
{1+2a_R\bar p_\eta+|a(\eta A)|^2\bar p_\eta^2} \ , \\
\mbox{Im} (T_{\eta})&=&\frac{-\bar p_\eta a_I}
{1+2a_R\bar p_\eta+|a(\eta A)|^2\bar p_\eta^2} \ .
\end{eqnarray}
Thus, when moving from above the threshold to below the threshold
the real part and the imaginary part of the $\eta$-nucleus scattering
length interchange their roles. Because of that also the signs of
these two quantities enter in a different way. Since unitarity fixes
the sign of the imaginary part, i.e. $a_I\ge 0$, this feature
opens the unique opportunity
to access the sign of the real part of the $\eta$-nucleus scattering
length by measuring the energy dependence of the cross section
ratio Eq. (\ref{rdef}) around the $\eta$ threshold!

The only crux in this kind of analysis is the occurrence of the
phase $\phi$ which is unknown.
However, we will argue below that the knowledge of $\phi$ is not
necessary for the analysis we propose, i.e.  we will show
that different signs of ${\rm Re}\, a(\eta A)$ lead to qualitatively
different results for the energy dependence of the cross section
ratio $R$ so that the two cases can be distinguished experimentally
even without knowledge of $\phi$.

 As should be clear from  Eq. (\ref{re}) 
 a variation in $\phi$ does not introduce any peculiarities but
leads to a rather smooth behaviour of $\mbox{Re}
({\cal M}_{\eta}{\cal M}_{\tilde \pi^0}^*)$.
Therefore, we look only at the dependence of the
ratio $R$ on the $\eta$ momentum for fixed values of $\phi$.
Thereby, we consider basically the whole range of $\phi$.
However, we restrict ourselves to those values of $\phi$ where
$R$ is smaller than two above the $\eta$ threshold, as is suggested
by the preliminary data from GEM \cite{machner2,machner3}.

As was mentioned above, the phase $\phi$ should depend on the pion
emission angle. Thus, any possible systematic error introduced
by the $\phi$ dependence could be explored and eliminated
by performing the measurement of the energy dependence of $R$
for a variety of pion angles.

Finally, for the $\eta\,^3{\rm He}$ scattering length we use the
values for the real and imaginary part as given in Eqs. (\ref{aetaex}),
which where extracted from the data in Ref. \cite{etadata}, and
we investigate the influence of different choices for
the sign of the real part of $a(\eta$$^3$He$)$ on $R$.
We should mention at this point, however, that the values of the real
and imaginary parts of of $a(\eta$$^3$He$)$ cannot be independently determined
by a fit to the $pd\to\, ^3{\rm He}\eta$ cross section based on
Eq. (\ref{tmat}). Rather, there is a correlation between them
with the consequence that all values fulfilling the relation
(units in $fm$)
\begin{equation}
a^2_R + 0.449 a^2_I + 4.509 a_I = 21.44
\label{Wil}
\end{equation}
lead to basically the same $\chi^2$ minimum \cite{wilkin2}.
In order to explore also the influence of this uncertainty
we employ several values for the $\eta\,^3{\rm He}$ scattering length.
I. e. we make the (certainly extreme) assumption that
$a_I= 0.5$ fm (the lowest limit for the imaginary part
in Eq. (\ref{aetaex})), which then leads to $|a_R|$ = 4.3 fm (c.f. Eq. (\ref{Wil})),
and we look at the other extreme as well by choosing the largest possible
value for $a_I$ which is still compatible with the data in
Ref. \cite{etadata} (see Eq. (\ref{aetaex})). Here we get $|a_R|$ = 2.4 fm, $a_I$ = 2.7 fm.
Of course, we also employ the central values of Eq. (\ref{aetaex}).

The other parameter values used in our analysis are:
$\theta_m$= 0.0015 \cite{theta};
$|{\cal M}_{\tilde \pi^0}|^2$ = 0.06 [$\mu b /sr$], which is extracted from
the $pd \to ^3\!{\rm \!He}\pi^0$ amplitude at a proton-pion relative angle
of $\theta_{p-\pi}$ = 180$^o$ and at energies around 
the $^3{\rm He}\eta$ threshold \cite{machner}.
The value of ${\cal M}^0_{\eta}$ depends on the employed $\eta\,^3{\rm He}$
scattering length. Here we obtained
$|{\cal M}^0_{\eta}|^2$ = 1.51 [$\mu b /sr$] (for $a_R+ia_I$ = $\pm 4.3 + i 0.5$ fm),
$|{\cal M}^0_{\eta}|^2$ = 1.74 [$\mu b /sr$] (for $\pm 3.8 + i 1.6$ fm), and
$|{\cal M}^0_{\eta}|^2$ = 1.93 [$\mu b /sr$] (for $\pm 2.4 + i 2.7$ fm),
respectively, by fitting to the $pd\to\, ^3{\rm He}\eta$ cross section
data \cite{etadata}.

\begin{figure}[h]
\begin{center}
\epsfig{file=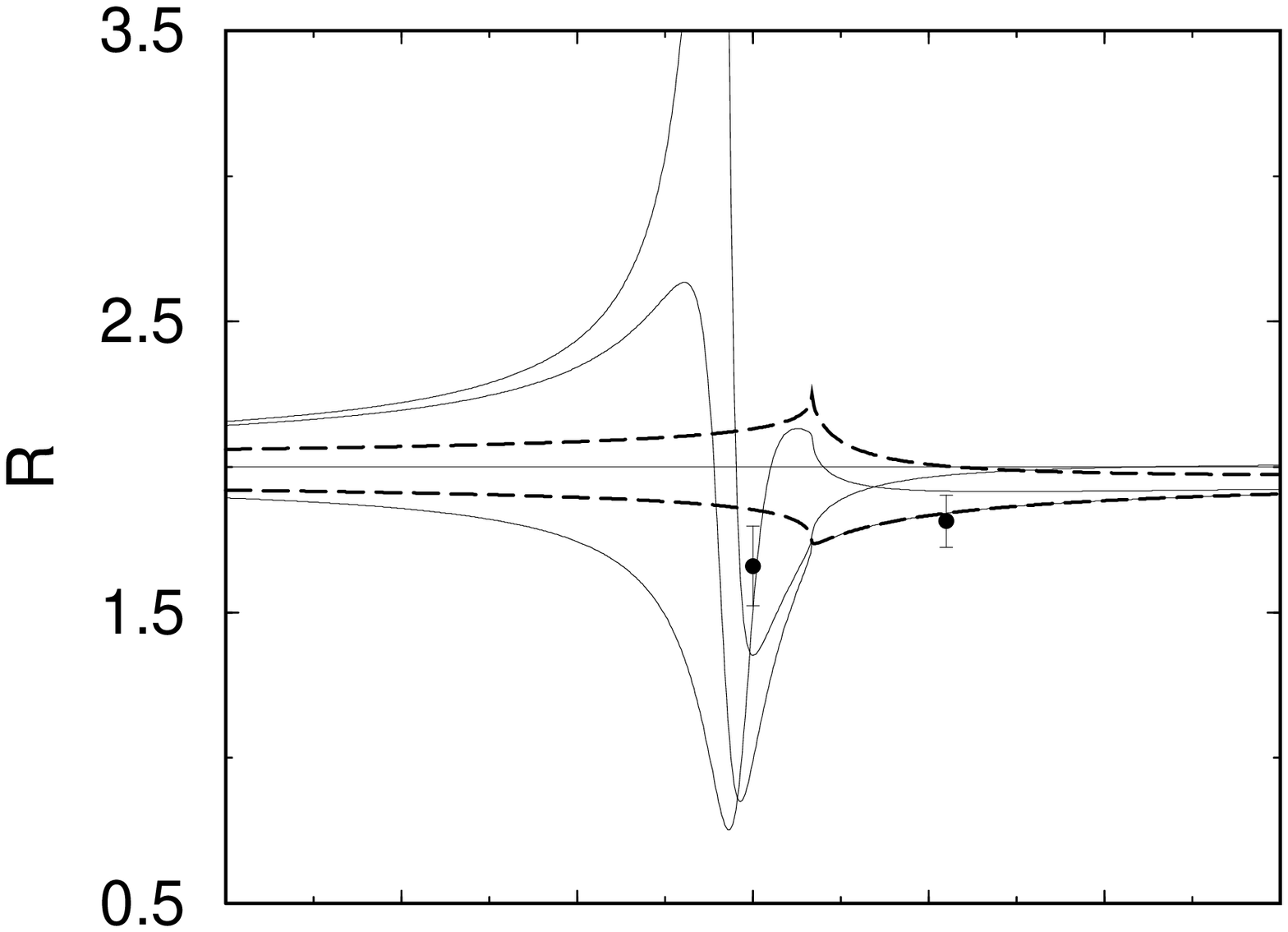,width=5.2cm}
\epsfig{file=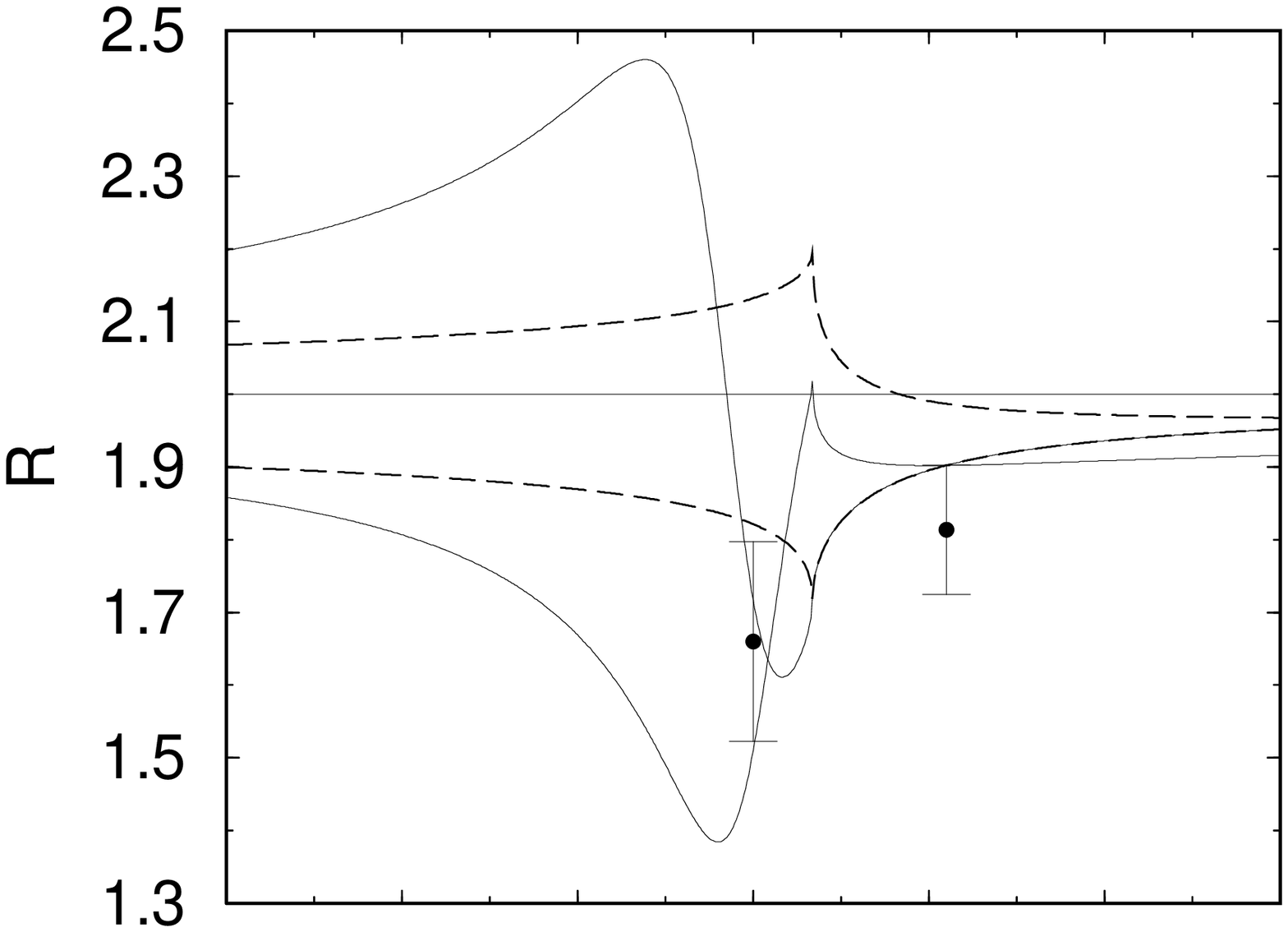,width=5.2cm}
\hspace*{0.27cm}\epsfig{file=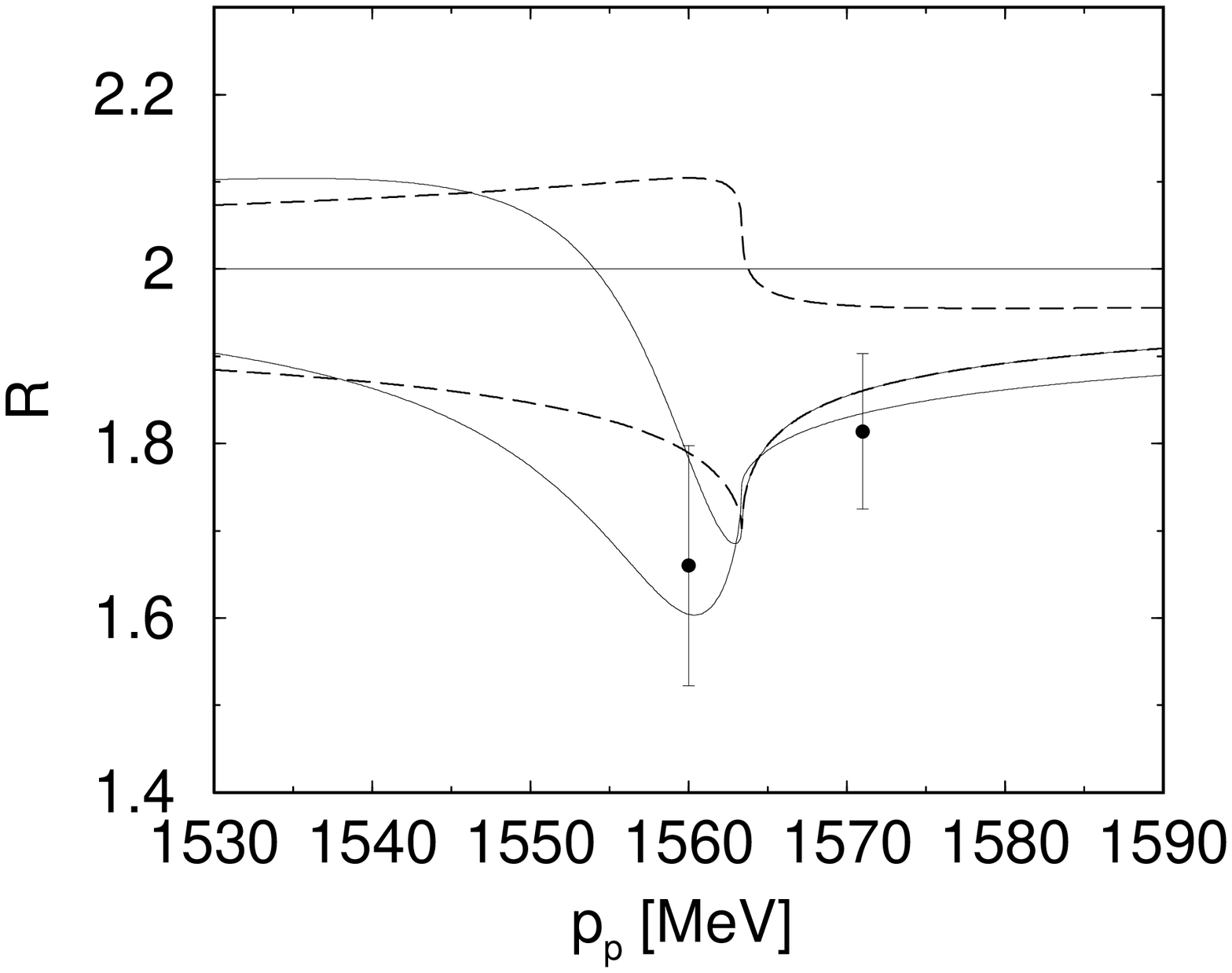,width=5.5cm}
\caption{Predictions for the cross section ratio $R$ for different
values and different signs of Re$(a(\eta\, ^3{\rm He}))$.
The $\eta\, ^3{\rm He}$ scattering length is ($\pm 4.3+i0.5$)fm
(upper panel),
($\pm 3.8+i1.6$)fm (middle panel), ($\pm 2.4+i2.7$)fm (bottom panel).
The curves are for individually selected values of the phase $\phi$,
cf. discussion in the text, where
the dashed lines correspond to a positive real part of the scattering length
and the solid lines correspond to a negative real part.
The horizontal solid line indicates the value of 2 for the ratio predicted by
isospin symmetry.
Note that the  scale is different for different panels.}
\label{ratio}
\end{center}
\end{figure}

The results of our investigation are presented in Fig. \ref{ratio}.
Though we have explored basically the whole available parameter
space we would like to concentrate here on a few but
exemplary cases. Varying the phase $\phi$ we found
examples where there is a very pronounced difference in the
energy dependence of the cross section ratio $R$ for the two
choices of the sign of $a_R$ and which, therefore, can be
easily distinguished in an experiment. On the other hand,
there are also cases where the differences in the results
around the $\eta$ production threshold can be very small.
Representative results for those ``best'' or ``worst'' cases
are shown in Fig. \ref{ratio}.
It may be noted that all results for positive values of $a_R$ are
basically between the dashed curves and for negative ones between the
solid curves, respectively, for any choice of $\phi$. But one should
keep in mind that the bounds alone are not that important.
The variation of the ratio $R$ with energy is the main
criterion for distinguishing between a positive or negative $a_R$
based on experimental data.

As a more qualitative feature we see that
for $a_R$ larger than zero (dashed lines) there is, in general,
a cusp-like structure of the ratio $R$ at the
$\eta$ production threshold (the corresponding proton momentum is
$p_p \approx 1563 $ MeV), whereas for $a_R$ smaller than zero
(solid lines) one observes a so-called rounded step \cite{cusp}.
Consequently, in the former case $|R-2|$ decreases more or less
monotoneously below the $\eta$ production threshold. On the
other hand,
for $a_R$ smaller than zero, $|R-2|$ increases and,
moreover, shows a strong momentum dependence.
For instance, in the upper and middle panels of Fig. \ref{ratio} one can see
that the curves with $a_R < 0$ have either a clear bump or a dip (or even both)
for some specific momenta below the threshold which can be easily
distinguished from the monotonously decreasing curves corresponding to $a_R > 0$.

A detailed inspection of Fig. \ref{ratio} reveals that
in some cases it is insufficient to determine the cross section ratio
only in a small energy range around the $\eta$-threshold
-- despite
the fact that the momentum dependence is strikingly different for
different signs of $a_R$ for all values of the angle $\phi$.
Such a situation can be seen in the middle panel of Fig. \ref{ratio}.
Here one of the sample results for $a_R < 0$ (solid curve) exhibits a dip
very close to the threshold which would be difficult to distinguish from the
cusp-like structure of similar magnitude produced by a calculation
using $a_R > 0$ (dashed curve) -- given the present accuracy of the
experimental data -- if one looks only into a very narrow energy range.
Here measurements over a wider energy range are necessary.
It is obvious from this figure that measurements at 5-20 MeV/c below
the threshold will allow to distinguish the different scenarios.

We also observed some cases where seemingly only a rather high
experimental accuracy would allow to distinguish between the two
scenarios. An example for this can be found in the lower panel of
Fig. \ref{ratio}. Here we see a sample result with $a_R < 0$ where
the dip is still fairly close to the threshold and where also the
momentum dependence of $R$ below the threshold is similar to the
one produced by a corresponding calculation based on $a_R > 0$.

In this context, let us emphasize, however, that increasing the
experimental accuracy is not the only option one has.
Further measurements performed at different angles
between proton and pion should be also helpful, since then the
phase $\phi$ is changed as well and could be shifted to a
different range of values where a discrimination
between the two signs for $a_R$ is much better feasible.

Nevertheless, it is obvious that the possibility to distinguish between
the two scenarios depends to a certain extent on the magnitude of $|a_R|$
and the differences in the cross section ratio caused by a positive or
negative sign are getting smaller with decreasing value
of $|a_R|$. As we discussed above and as can be seen in the
lower panel of Fig. \ref{ratio}, already in the case of
$a(\eta ^3{\rm He})=(\pm 2.4+i2.7)$fm it is somewhat tricky
to discriminate between the two signs for $a_R$
and the situation will be even more involved should $|a_R|$ be
still smaller.

But even in such a situation interesting conclusions can be drawn from
the cross section ratio.
In order to understand that we need to remind the reader
that in case of a complex scattering
length the condition $a_R < 0$ alone is not sufficient for having a
bound state. Here there is an additional constraint,
namely that $|a_I| < |a_R|$ \cite{HaLi2}. The results presented
above indicate that the possibility to distinguish between the two
scenarios for the sign of $a_R$ is getting more and more difficult just in
such cases where this constraint is not fulfilled anymore. Therefore, even
if the measured cross section ratio shows features like those in
the lower panel of Fig. \ref{ratio} --  which would make it difficult
if not impossible to determine the sign of $a_R$ -- it would still
allow to rule out a bound state.

To summarize, we have demonstrated that charge symmetry breaking
can be used as a tool to get direct access to the real part of the
$\eta$-nucleus scattering length and specifically to its sign.
The knowledge of this sign is important for drawing conclusions
about the possible existence of $\eta$-nucleus bound states.
In the present paper we outlined the general idea and strategy
for a corresponding analysis
and we exemplified its feasibility for the reactions
$pd\to\, ^3{\rm H}\pi/^3{\rm He}\pi$.
With the same initial state one can also look at $NN\to d\pi$
with one nucleon being a spectator. Again,
in the case of charge symmetry $R$ defined analogously to Eq.~(\ref{rdef0})
will be 2.  However, close to the $\eta$ threshold
a significant deviation from this value should be observed
allowing one to determine the sign of Re($a(\eta d)$). In the same way,
bombarding a $^3{\rm H}$ target with protons allows access to
Re($a(\eta \alpha)$) and so on. All these experiments are presently 
feasible, e.g.,  at the CELSIUS as well as COSY accelerators.
In addition to the pion cross section ratio, of course, the corresponding
$\eta$ cross section should be measured to high accuracy.
Only a profound knowledge of the energy dependence of the $\eta$ 
cross section allows to sufficiently
constrain the magnitudes of the relevant $\eta$-nucleus scattering lengths so that an
analysis along the lines suggested becomes practicable.

\vskip 0.5cm

The authors would like to thank Barry R. Holstein for useful discussions.
Financial support for this work was provided in part by the
international exchange program between DAAD (Germany, Project No.
313-SF-PPP-8) and the Academy of Finland (Project No. 41926).


\begin{references}
%
\bibitem{HaLi} Q. Haider and L.C. Liu,
Phys. Lett. B {\bf 172}, 257 (1986); {\bf 174}, 465 (E) (1986).
%
\bibitem{Bahle}
R.S. Bhalerao and L.C. Liu, Phys. Rev. Lett. {\bf 54}, 865 (1985).
%
\bibitem{x1} T. Ueda, Phys. Rev. Lett. {\bf 66}, 297 (1991);
N.V. Shevchenko {\it et al.}, Eur. Phys. J. A {\bf 9}, 143 (2000);
S. Wycech and A.M. Green, Phys. Rev. {\bf C 64}, 045206 (2001).
%
\bibitem{x2} A. Fix and H. Arenh\"ovel, Eur. Phys. J. A {\bf 9}, 119 (2000);
H. Garcilazo and M.T. Pe\~na, Phys. Rev. {\bf C 63}, 021001 (2001).
%
\bibitem{wycech} S. Wycech, A.M. Green, and J.A. Niskanen,
Phys. Rev. {\bf C 52}, 544 (1995).
%
\bibitem{willis}
N. Willis {\it et al.}, Phys. Lett. B {\bf 406}, 14 (1997).
%
\bibitem{HaLi2} Q. Haider and L.C. Liu,
Phys. Rev. {\bf C 66}, 045208 (2002).
%
\bibitem{newton} R.G. Newton,
{\it Scattering Theory of Waves and Particles}
(Springer-Verlag, New York, 1982).
%
\bibitem{watson}
M. Goldberger and K.M. Watson, {\it Collision Theory} (Wiley, New York, 1964).
%
\bibitem{wilkin}
C. Wilkin, Phys. Lett. B {\bf 331}, 276 (1994).
%
\bibitem{machner}
A.~Magiera and H.~Machner,
Nucl.\ Phys.\ A {\bf 674}, 515 (2000).
%
\bibitem{machner2}
H.~Machner, Proceedings of the International School of Nuclear Physics,
``Quarks in Hadrons and Nuclei'', Erice, Italy, Sept. 16-24, 2002,
Prog. in Part. and Nucl. Phys. {\bf 50}, in print.
%
\bibitem{machner3}
A. Magiera et al.,
Meson 2002. Proceedings of the 7th
International Workshop on Production, Properties,
and Interaction of Mesons,
edited by L. Jarczyk {\it et al.},
(World Scientific, Singapore, 2003), p. 155.
%
\bibitem{wycech2} A.M. Green and S. Wycech,
Meson 2002. Proceedings of the 7th
International Workshop on Production, Properties,
and Interaction of Mesons,
edited by L. Jarczyk {\it et al.},
(World Scientific, Singapore, 2003), p. 323.
%
\bibitem{etadata}
B. Mayer {\it et al.}, Phys. Rev. {\bf C 53}, 2068 (1996).
%
\bibitem{wilkin2}
C. Wilkin, Phys. Rev. {\bf C 47}, 938 (1993).
%
\bibitem{theta}
Ll. Ametller, C. Ayala, and A. Bramon, Phys. Rev. {\bf D 30}, 674 (1984).
%
\bibitem{cusp} R.G. Newton, Ann. Phys. (N. Y.) {\bf 4}, 29 (1958).
%
\end{references}
\end{document}